\documentstyle[mymulticol,pra,aps,amsfonts]{revtex}

\newcommand{\cL}{{\cal L}}

\newcommand{\cF}{{\cal F}}

\newcommand{\cM}{{\cal M}}
\newcommand{\cA}{{\cal A}}
\newcommand{\cG}{{\cal G}}

\newcommand{\beq}{\begin{equation}}
\newcommand{\eeq}{\end{equation}}
\newcommand{\beqy}{\begin{eqnarray}}
\newcommand{\eeqy}{\end{eqnarray}}

\makeatletter
\def\@beginTheorem#1#2{\trivlist \item[\hskip \labelsep{\bf #1\ #2}]}
\def\@opargbegintheorem#1#2#3{ \trivlist
      \item[\hskip \labelsep{\bf #1\ #2\ (#3)}]}
\makeatother

\makeatletter
\def\@beginLemma#1#2{\trivlist \item[\hskip \labelsep{\bf #1\ #2}]}
\def\@opargbeginLemma#1#2#3{ \trivlist
      \item[\hski
 \labelsep{\bf #1\ #2\ (#3)}]}
\makeatother

\makeatletter
\def\@beginDefinition#1#2{\trivlist \item[\hskip \labelsep{\bf #1\ #2}]}
\def\@opargbeginDefinition#1#2#3{ \trivlist
      \item[\hskip \labelsep{\bf #1\ #2\ (#3)}]}
\makeatother

\makeatletter
\def\@beginCorollary#1#2{\trivlist \item[\hskip \labelsep{\bf #1\ #2}]}
\def\@opargbeginCorollary#1#2#3{ \trivlist
      \item[\hskip \labelsep{\bf #1\ #2\ (#3)}]}
\makeatother

\makeatletter
\def\@beginExample#1#2{\trivlist \item[\hskip \labelsep{\bf #1\ #2}]}
\def\@opargbeginExample#1#2#3{ \trivlist
      \item[\hskip \labelsep{\bf #1\ #2\ (#3)}]}
\makeatother

\def\C{{\mathbb{C}}}

\def\R{{\mathbb{R}}}

\def\N{{\mathbb{N}}}

\newcommand{\cH}{{\cal H}}

\newcommand{\Einr}{\hspace{1cm}}

\title{Distinguishing $n$ Hamiltonians on $\C^n$ by a single measurement}

\author{D. Janzing\thanks{Electronic address: janzing@ira.uka.de}  and 
Th. Beth}
\address{Institut f\"ur Algorithmen und Kognitive Systeme, Am Fasanengarten 3a,
    D--76\,131 Karlsruhe, Germany}

\begin{document}
\maketitle

\begin{abstract}
If an experimentalist wants to decide which one of $n$ possible
Hamiltonians acting on an $n$ dimensional Hilbert space is present,
he can conjugate the time evolution by an appropriate sequence
of known unitary transformations
in such a way  that the different Hamiltonians result in mutual
orthogonal
final states. We
present a general scheme providing such a sequence.
\end{abstract}

\begin{multicols}{2}

Controling simple quantum systems has become 
a large field of research during the last decade.
Experimental and theoretical investigations deal with
the preparation of certain quantum states, the implementation
of unitary transformations and the design of measurement procedures 
for different quantum  observables.
Whereas the problem of optimal information gain
about unknown quantum {\it states} is a large field of research
(see e.g. \cite{Fu96}), discussions about optimal estimation of
unknown quantum {\it evolutions} are comparably rare.
A strategy for estimating an 
arbitrary
unknown unitary transformation
is developed in \cite{AJV00} 
and  in \cite{AP00} for
general quantum operations.
Those approaches assume that the dynamical evolution to be 
estimated is taken from
an infinite set of possibilities.
The problem of estimating an unknown Hamiltonian can arise
in various contextes:
Assume we want to use a single quantum system
in order to detect classical fields, e.g., a spin particle
as detector for a magnetic field.
We expose the test particle to the field for a certain time period
and 
estimate the field strength by measuring the particle's quantum state.
Assume the experimentalist is allowed to perform arbitrary 
unitary transformations
on the test particle, expose the particle to the field again,
repeat this several times
and perform a single measurement at the end.
What is the best procedure for estimating the field?
If the set of possible values for the field strength 
is larger than $n$, a single measurement of the test particle can
only allow {\it estimations} of the field.
By basic quantum mechanics, it is well-known that a set of {\it states}
is perfectly distinguishable by a single measurement
if and only if their density matrices have disjoint support.
Led by this simple statement concerning the distinguishability
of {\it states}, 
we focus on the question of distinguishing 
between $n$ possible 
Hamiltonians  $\{H_1,\dots, H_n\}$
of a quantum system on the Hilbert space $\cH:=\C^n$
and show that they are always perfectly distinguishable
provided they do not only differ by an {\it additive} constant.
We assume that the experimentalist is allowed to
prepare the initial state, to perform definite
unitary transformations interrupting the unknown natural evolution
and to perform an arbitrary measurement at the end. 
The assumption about the restricted set of possibilities 
is more natural than it might seem at first sight. 
Take the following model of a measurement interaction 
(compare \cite{Ki95}):
on the joint Hilbert space of the measured system and the measurement 
apparatus we assume to have the Hamiltonian
\[
G:=\sum_{j} P_j\otimes H_j,
\]
where $(P_j)_j$ is the family of spectral projections of the measured observable
and $H_j$ are different self-adjoint operators moving
the pointer of the measurement apparatus conditioned on the state
of the measured system. Assume that we do not have any direct access
to the measured system and that we are not able to change the interaction at
all. The only way to use the interaction for a measurement procedure
consists in 
initializing the measuring device,
waiting (i.e. implementing $e^{iGt}$)
and interrupting this evolution
several times by implementing local unitary transformations on the
measurement apparatus in order to get mutual orthogonal pointer states
for different Hamiltonians $H_j$.

Our considerations 
show that this is always possible  (if $H_j -tr(H_j)\neq H_i -tr(H_i)$ for 
$i\neq j$) 
and give a general rule for such a
quantum algorithm.

The algorithm consists of quite a large number of steps;
since we are working in the Lie algebra instead of the Lie group
our scheme requires arbitrarily many unitary transformations
(close to the identity) in order to
obtain the correct result with arbitrary reliability.
We are convinced that there exist much simpler algorithms
for particular sets of $n$ Hamiltonians.
Whether or not there are general rules requiring only a few steps
is unclear. Developing short procedures for the general case 
might result in computationally
hard word problems
in the Lie group $SU_n$, whereas our classical precomputation 
consists only  in solving linear equations for the price of obtaining
only approximative solutions.

Firstly we  present an example of $n$ Hamiltonians 
which can be distinguished easily:
Set $H_j:= j D$ with $D:=diag (1,2,\dots,n)$.
By waiting the time $t=2\pi/n$ we have implemented the unitary transformations
$e^{ij2\pi D/n}$. Take the initial vector
$|\psi\rangle:=(1,1,\dots,1)^T$. Then the states $e^{ij2\pi D /n}|\psi\rangle$
are orthogonal for different values of $j$  since they are the discrete 
Fourier transforms of the canonical basis vectors of $\C^n$.
In the rest of the paper we show that the general problem can
be reduced to this example. For doing so we start by developing
some technical tools: 

By waiting the time $t$, we have implemented
the transformation $e^{iHt}$  for the unknown Hamiltonian
$H\in \{H_1,\dots, H_n\}$.
We show that there is a procedure
simulating $e^{-iHs}$ for arbitrary $s$:
Choose a finite subgroup $\cG$ of $SU_n$ acting irreducibly
on $\cH$. Then 
\[
\sum_{U\in \cG} UHU^\dagger
\]
is an operator commuting with
every $U\in \cG$ and is therefore a multiple of the identity operator
by Schur's Lemma (this fact is used in decoupling techniques \cite{VKL98}).
Without loss of generality we assume every $H_j$ to be traceless.
Then one has $\sum_{U\in \cG} U HU^\dagger=0$ and hence
$\sum_{U\in \cG\setminus \{1\}} UHU^\dagger=-H$.
We obtain
\[
\lim_{m\to\infty}(\Pi_{U\in \cG\setminus \{1\}}Ue^{iHt/m}U^\dagger)^m=e^{-iHt}.
\]
Set $G:=\{1,U_1,\dots,U_l\}$.
Then for
large $m$ we have approximately an implementation of $e^{-iHt}$
as follows:

\vspace{0.5cm}

\noindent  
{\tt begin}

\begin{quote}
\noindent
{\tt for $k=1$ to $m$ do}

\noindent
\Einr {\tt for $s=1$ to $l$ do}

\noindent
\Einr \Einr {\tt implement $U_s$}

\noindent
\Einr \Einr {\tt wait the time $t/m$}

\noindent
\Einr \Einr {\tt implement $U_s^\dagger$}

\end{quote}
\noindent
{\tt end}.

\vspace{0.5cm}

The possibility of implementing $e^{iHt}$ even for negative $t$ 
is decisive for using Lie algebraic tools in the sequel:
Let $\cA$ be the Lie algebra of traceless self-adjoint operators
acting on $\cH$. 

By using the well-known formula 
\[
\lim_{m\to\infty}(e^{iH/m}e^{iA/m}e^{-iH/m}e^{-iA/m})^{m^2}=e^{i[H,A]}
\]
we can design an algorithm simulating the unitary
\[
e^{-[H,A]s}
\]
for arbitrary $s\in \R, A\in \cA$ with arbitrary small error. 
In the same way we conclude more generally:

\vspace{0.5cm}
\noindent
{\bf Lemma 1} 
Let $\cF,\cG:\cA\rightarrow \cA$ be
arbitrary (not necessarily linear) functions.  
Assume  there exist for every $s\in \R$  procedures
for simulating the unitary transformations 
\[
e^{i\cF(H)s} 
\]
and for simulating
\[
e^{i\cG(H)s}
\]
with arbitrary small error
for  the unknown  Hamiltonian $H\in \{H_1,\dots,H_n\}$.
Then there are procedures simulating
\[
e^{i[\cG(H),\cF(H)]s}
\]
and 
\[
e^{i[\cF(H),A]s}
\]
for every $A\in \cA$ and
every $s\in \R$ with arbitrary small error.

\vspace{0.5cm}

\noindent
Obviously, for every $A\in \cA$ we can find an algorithm
performing
$i[H,A]=:Ad(H) (A)$. Hence we can find for
every $k\in \N$ an algorithm performing $(Ad(H))^k(A)$.
We conclude:

\vspace{0.5cm}
\noindent
{\bf Corollary}
Let $\cF:\cA\rightarrow \cA$ be an arbitrary function.
Assume that for every required accuracy
and every $s\in \R$ 
there exists a procedure
such that
\[
e^{i\cF(H)s}
\]
is implemented.
Then Lemma 1 provides a scheme for implementing
\[
e^{ip(Ad(\cF(H))))(A)}
\]
where $p$ is an arbitrary real polynomial
and $A\in \cA$.
\vspace{0.5cm}

Furthermore we will need the following Lie algebraic Lemma:

\vspace{0.5cm}
\noindent
{\bf Lemma 2}
Let $Hom(\cA,\cA)$ be the ring of $\R$-linear maps
on the real vector space $\cA$.
Then there is no proper subring of $Hom(\cA,\cA)$ containing all the maps
of the form $B\mapsto i[A,B]$ with arbitrary $A\in \cA$.

\vspace{0.5cm}

\noindent
{\it Proof}
Define $D_1:=diag(1,-1,0,\dots,0)$,
$D_2:=(0,1,-1,0,\dots,0)$,\dots, $D_{n-1}:=(0,\dots, 0,1,-1)$.
Furthermore let $X_{jk}$ for every unordered pair $(j,k)$ with $j<k$
be the
matrix with $1$ at the positions $j,k$ and $k,j$ and zero elsewhere.
Let $Y_{j,k}$ be defined in an analogue way with entries $i$ and $-i$ at
positions $j,k$ and $k,j$ respectively.
The set of these $n^2-1$ matrices forms a basis of the vector space $\cA$.
Since all the basis vectors are unitarily equivalent (note the analogy
to the Pauli-matrices),
there always exists a map in the ring generated by maps of the form $i[A,.]$
mapping one basis vector on the other.
In order to show, that every map $\cL\in Hom(\cA,\cA)$
can be obtained  by sums and concatenations of  maps $i[A,.]$ it is 
therefore
sufficient to prove that a map $\cM$ can be generated with
the following two properties: (1) The kernel of $\cM$ contains every 
 basis vector except $D_1$ and (2)
$\cM(D_1)$ is  proportional to $Y_{1,2}$.
Choose a finite subgroup $S$ of $SU_n$ 
acting trivially on the vector space spanned by the first basis
vector $|1\rangle$ and irreducibly on its orthogonal complement 
$|1\rangle^\perp$.
The sub ring we are looking for
contains clearly the map
$\cL(A):=(1/|S|)\sum_j U_j A U_j^\dagger$ since $U_jA U_j=e^{Ad(B_j)}(A)$ if
$B_j$ is chosen such that
$e^{iB_j}=U_j$.
Due to Shur's Lemma every operator in the image of $\cL$  
is a multiple of the identity on $|1\rangle^\perp$.
Since the trace on the subspace $|1\rangle^\perp$ 
is invariant, the restriction of $\cL$ to  $|1\rangle^\perp$
is given by $B\mapsto tr(B)1$. 
Explicitly one obtains
\[
\cL(A)= (1-P)A(1-P) + tr(PAP)P,
\]
where $P$ is the projector onto $|1\rangle^\perp$.
Clearly $\cL$ annihilates all  the basis vectors except $D_{1}$.
Define $\cM$ by $\cM(B):= i[X_{1,2},\cL(B)]$
Easy calculation shows  that $\cM(D_1)$ is proportional to $Y_{1,2}$.
$\Box$

\vspace{0.5cm}

\noindent
Since $i[A,B]=\lim_{s\to 0} (e^{iAs} B e^{-iAs} - B)/s$
we can conclude that $i[A,.]$ is an element of the vector space
spanned by the maps $(B\mapsto e^{iAs} B e^{-iAs})_{s\in \R}$.
Hence we obtain:

\vspace{0.5cm}
\noindent
{\bf Corollary}
Let $\cL\in Hom(\cA,\cA)$. Then there is a set of unitaries $U_1,\dots,U_m$
and real numbers $c_1,\dots,c_m$
such that $\cL(B)=\sum_{j\leq m} c_j U_{j} B U_j^\dagger$.
\vspace{0.5cm}

In order to obtain a constructive statement one can either take the 
approximative solution defined by the limit above or one can 
write $Ad(A)$ as a finite linear combination of maps 
$(e^{Ad(A)s})_{s\in \R}$ by
solving the corresponding equation for the eigenvalues
of $Ad(A)$.
We conclude:

\vspace{0.5cm}

\noindent
{\bf Lemma 3}
Let $\cF: \cA\rightarrow \cA$ be arbitrary. 
If there is a scheme implementing $e^{i\cF(H)s}$ 
for the unknown Hamiltonian $H\in \{H_1,\dots,H_n\}$ then
\[
e^{i \cL(\cF(H))s}
\]
for arbitrary $\cL\in Hom(\cA,\cA)$ can be implemented
with arbitrary small error 
by
\[
(U_1 e^{i\cF(H)c_1s/k} U_1^\dagger U_2 e^{i\cF(H)c_2s/k}U_2^\dagger \dots 
U_me^{i\cF(H)c_m s/k}
U_m^\dagger)^k
\]
where $U_j$ are the unitaries and $c_j$ are the coefficients
corresponding
to $\cL$ in the sense of the corollary to Lemma 2 and
$k$ is large enough to keep the error small.

\vspace{0.5cm}

\noindent
Now we are able to construct our algorithm:
Choose an operator $G\in \cA$ with exactly two different eigenvalues, called
$\alpha$ and $\beta$.
Choose $\cL\in Hom(\cA,\cA)$ in such a way that
$\cL(H_j)=\lambda_j G$ with $\lambda_j > 0$ and  $\lambda_i\neq \lambda_j$.
This is possible due to
basic linear algebra.
The map $Ad(G):=i[G,.]$
has the eigenvalues $\pm i(\alpha-\beta)$  and $0$.
The spectrum of the map $Ad(\lambda_j G)$ is hence given by the values
$\pm \lambda_j i (\alpha-\beta),0$. Choose a real 
polynomial $p$ such that
$p(\pm \lambda_j i (\alpha-\beta))=\pm j i (\alpha-\beta)$ and $p(0)=0$. 
Due to the functional calculus for the diagonalizable operator
$Ad(G)$ 
this implies
\[
p( \lambda_j ( Ad(G)))= j Ad(G)
\]
By defining $C:=Ad(G)(A)$
for arbitrary $A\in \cA\setminus \{0\}$ we obtain
\begin{equation}\label{AdG}
p(Ad(\lambda_j G)) (A)= j C. 
\end{equation}

Now choose a map $\tilde{\cL}\in Hom(\cA,\cA)$ such that
\[
\tilde{\cL}(C)=D 2\pi/n.
\]
We obtain $\tilde{\cL}(p(Ad( \cL(H)))(A))=jD 2\pi/n$.

The classical precomputation for our
algorithm can be sketched as follows:
 
\begin{enumerate}
\item Choose an element $G\in \cA$ with two-valued spectrum and
      find a linear map $\cL$ such that $\cL(H_j)=\lambda_j G$ with different 
      values $\lambda_j$.

\item Find a set of unitary transformations $U_1,\dots,U_l$ 
      and a set of real numbers $c_j$ such that
      $\cL(B)=\sum_j c_j U_j B U_j^\dagger$ for every $B\in \cA$.
      This is possible 
      due to the corollary
      to Lemma 2.  
               
\item Choose a polynomial $p$ such that $p(\pm \lambda_j i(\alpha-\beta))
      =\pm j$ and $p(0)=0$,
      if $\alpha,\beta$ are the eigenvalues of $G$.

\item Choose an arbitrary  operator $A\in \cA\setminus \{0\}$ and a
      map $\tilde{\cL}$ such that $\tilde{\cL} \big(p(Ad(G))(A)\big)=
      D 2\pi/n$.
      Find a set of unitary operators $V_1,\dots, V_m$ and real numbers
      $d_j$  such that
      $\tilde{\cL} (B)=\sum_j d_j V_j B V_j^\dagger$. 

\end{enumerate}

Now we sketch the required sequence of quantum operations
as follows:

\begin{enumerate}
\item
Prepare the initial state $|\psi\rangle:=(1/\sqrt{n})(1,\dots,1)^T$.
\item
Call a subroutine performing
the evolution $e^{ijD2\pi/n}$ if the Hamiltonian $H_j$ is present.
\item Measure in the basis defined by the discrete
Fourier transforms of the canonical basis vectors of $\C^n$.
If the result is the $j^{th}$ basis state then the Hamiltonian
$H_j$ is present.
\end{enumerate}

The subroutine called in step (2) is recursively defined:
The implementation of 
\[
e^{ijD2\pi/n}=e^{ij \tilde{\cL}\big(p(Ad(L(H_j)))(A)\big)}
\]
is based on Lemma 2
by calling a subroutine simulating
\[
e^{ip(Ad(\cL(H_j)))(A)s}
\]
for small $s$ several times.
The implementation of the latter is based
on the  corollary to Lemma 2 by calling a subroutine 
for implementing
\[
e^{i\cL(H_j)s}
\]
several times (Lemma 3). 

\end{multicols}

\section*{Acknowledgements}
We thank P. Wocjan for useful discussions and important corrections.
This work has been supported by grants of the project Q-ACTA of the European
Union.

\end{document}